# The effects of mixedness and entanglement on the properties of the entropic uncertainty in Heisenberg model with Dzyaloshinski-Moriya interaction

Xiao Zheng(郑晓)[1], Guo-Feng Zhang(张国锋)[1,2,*]

[1]*Key Laboratory of Micro-Nano Measurement-Manipulation and Physics (Ministry of Education), School of Physics and Nuclear Energy Engineering, Beihang University, Xueyuan Road No. 37, Beijing 100191, China*
[2]*State Key Laboratory of Low-Dimensional Quantum Physics, Tsinghua University, Beijing 100084, China*

**Abstract:** The effects of mixedness and entanglement on the lower bound and tightness of the entropic uncertainty in the Heisenberg model with Dzyaloshinski-Moriya (DM) interaction have been investigated. It is found that the mixedness can reflect the essence of the entropic uncertainty better than the entanglement. Meanwhile, the uncertainty of measurement results will be reduced by the entanglement and improved by the mixedness. The entanglement can destroy the tightness of the uncertainty, while the tightness will be improved with the increasing of the mixedness. In addition, the tightness of the uncertainty in Heisenberg model can be expressed as a function of the magnetic properties, the strength of the DM interaction as well as the mixedness of the state and the functional form has no relationship with temperature. What's more, the entropic uncertain inequality becomes uncertain equality when the mixedness of the system reach the minimum value. For a given mixedness, the tightness will be reduced with the increasing of the strength of DM interaction at the antiferromagnetic case while the situation is just the opposite for the ferromagnetic case.

## I. INTRODUCTION

The uncertainty principle is one of the most remarkable characters of quantum mechanics as well as a fundamental departure from the principle of classical physics [1-12]. Any pair of incompatible observables complies with a certain form of uncertainty relationship, the constraint of which sets ultimate bound on the measurement precision for these quantities and provides a theoretical basis for new technologies, such as, quantum cryptography in quantum information [13]. The traditional uncertainty relations is presented in the form of the standard deviation $\Delta R \Delta S \geq 1/2 |[R,S]|$ for two arbitrary incompatible observables $R$ and $S$[1,14]. And another method to quantify the uncertainty for any two general observables is based on the entropic measures [15]. The initial version of entropic uncertainty relation was first given by Kraus[3] and then proved by Maassen and Uffink[16], which reads as

$$H(R) + H(S) \geq \log_2 \frac{1}{c} \quad (1)$$

where $H(Y)$ indicates the Shannon entropy of the probability distribution of the outcomes when $Y$ is measured($Y \in \{R, S\}$). $\log_2(1/c)$ is used to quantify the complementarity of $R$ and $S$, in which $c = \max_{r,s}|\langle \Psi_r | \Phi_s \rangle|^2$ for no degenerate observables, with $|\Psi_r\rangle$ and $|\Phi_s\rangle$ being the eigenvectors of $R$ and $S$.

While in recent, Berta *et al.* [5] showed that the uncertainty bound of the entropic uncertainty could actually be violated with the aid of a quantum memory. As shown in Fig.1, they imagine an uncertainty game between two players Alice and Bob, where Bob prepares two entangled particles and sends one to Alice, who then carries out one measurement(R or S) and announces which measurement her choose to Bob. We call the particle (A) which is sent to Alice as the measured system and remaining one (B) as memory system. The uncertainty of Bob, who has access to the memory system, about the result of measurements on measured system, is bounded by

$$H(R|B) + H(S|B) \geq \log_2 \frac{1}{c} + H(A|B), \quad (2)$$

where $H(A|B) = H(\rho_{AB}) - H(\rho_B)$ is the conditional von Neumann entropy of the density operator $\rho_{AB}$. On the left side of the inequality, $H(Y|B) = H(\rho_{YB}) - H(\rho_B)$, which is that of the post measurement state $\rho_{YB} = \sum_Y (|\Psi_Y\rangle\langle\Psi_Y| \otimes I)\rho_{AB}(|\Psi_Y\rangle\langle\Psi_Y| \otimes I)$, represents uncertainty of the measurement outcomes of Y conditioned on the prior information stored in B. Here $H(\rho) = -\text{tr}(\rho \log_2 \rho)$, $\rho_B = \text{tr}_A(\rho_{AB})$, $I$ stands for identical operator, and $|\Psi_Y\rangle$ is the eigenvectors of $Y$, $Y \in (R, S)$.

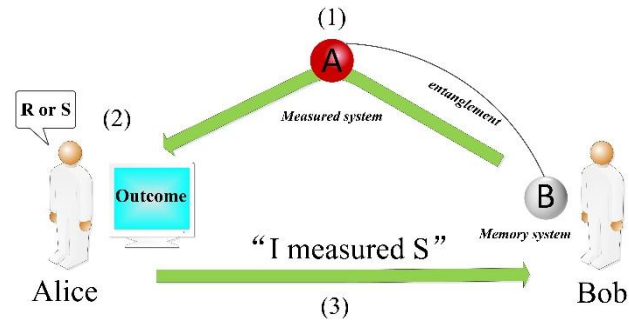

Fig.1: Illustration of the uncertainty game. (1) Bob prepares two entangled particles and sends one to Alice. (2)

Alice measures either *R* or *S* and notes her outcome. (3) Alice announces her measurement choice to Bob. Bob's goal is then to minimize his uncertainty about Alice's measurement outcome.

This new generalized entropic uncertainty principle has been recently confirmed experimentally [7], which has ignited the interest of people to investigate its potential applications from various aspects [5]. In this paper, we mainly research the effect of the mixedness and the entanglement on the properties of entropic uncertainty relation in the Heisenberg model with DM interaction. An outline of the reminder of the paper is as follows. In Sec. II, the physical model about two-qubits Heisenberg XX chain with DM interaction will be introduced. The properties of entropic uncertainty inequality in the Heisenberg system are studied in Sec. III. Finally, Sec. IV is devoted to the discussion and conclusion.

## II. Heisenberg Model with DM Interaction

In this paper, the Heisenberg model we investigated can be described by

$$H_{DM} = \frac{J}{2}[(\sigma_{1x}\sigma_{2x} + \sigma_{1y}\sigma_{2y} + \sigma_{1z}\sigma_{2z}) + \vec{D} \cdot (\vec{\sigma_1} \times \vec{\sigma_2})] , \qquad (3)$$

where $J$ is the real coupling coefficient and $\vec{D}$ is the DM coupling vector [17, 18]. The coupling constant $J > 0$ corresponds to the antiferromagnetic case and $J < 0$ to the ferromagnetic one. For simplicity, we choose $\vec{D} = D\vec{z}$. Then the Hamiltonian $H_{DM}$ becomes

$$H_{DM} = \frac{J}{2}[(\sigma_{1x}\sigma_{2x} + \sigma_{1y}\sigma_{2y} + \sigma_{1z}\sigma_{2z}) + D \cdot (\sigma_{1x}\sigma_{2y} - \sigma_{1y}\sigma_{2x})]$$

$$= J[(1 + iD)\sigma_{1+}\sigma_{2-} + (1 - iD)\sigma_{1-}\sigma_{2+}]. \qquad (4)$$

Without loss of generality, we denote the ground and excited state of a two-level particle by $|g\rangle$ and $|e\rangle$ respectively. The eigenvalues and eigenvectors of $H_{DM}$ are given by

$$H_{DM}|gg\rangle = \frac{J}{2}|gg\rangle$$

$$H_{DM}|ee\rangle = \frac{J}{2}|ee\rangle$$

$$H_{DM}|+\rangle = (J\sqrt{1 + D^2} - \frac{J}{2})|+\rangle$$

$$H_{DM}|-\rangle = (-J\sqrt{1 + D^2} - \frac{J}{2})|-\rangle , \qquad (5)$$

in which $|\pm\rangle = (1/\sqrt{2})(|ge\rangle \pm e^{i\theta}|eg\rangle)$ with $\theta = \arctan D$.

As thermal fluctuation is introduced into the system, the state of a typical solid state system at thermal equilibrium is $\rho(T) = (1/Z)\exp(-\beta H)$, where the Hamiltonian is denoted by $H$, $T$ represents the temperature and $Z = \text{Tr} \exp(-\beta H)$ is the partition function. Density matrix $\rho(T)$ based on the standard basis $\{|ee\rangle, |eg\rangle, |ge\rangle, |gg\rangle\}$ can be presented as

$$\rho(T) = \frac{1}{Z}\begin{pmatrix} \rho_{11} & 0 & 0 & 0 \\ 0 & \rho_{22} & \rho_{23} & 0 \\ 0 & \rho_{23}^* & \rho_{33} & 0 \\ 0 & 0 & 0 & \rho_{44} \end{pmatrix}, \tag{6}$$

with the elements

$$\rho_{11} = \rho_{44} = \exp(-\beta J/2), \tag{7}$$

$$\rho_{22} = \rho_{33} = \exp[\beta(J-\delta)/2](1+\exp(\beta\delta))/2, \tag{8}$$

$$\rho_{23} = \exp(i\theta)\exp[\beta(J-\delta)/2](1+\exp(\beta\delta))/2, \tag{9}$$

$$\rho_{23}^* = \exp(-i\theta)\exp[\beta(J-\delta)/2](1+\exp(\beta\delta))/2, \tag{10}$$

in which $Z = 2\exp(-\beta J/2)[1+\exp(\beta J)\cosh(\beta\delta/2)]$, $\beta = 1/kT$ and $\delta = 2J\sqrt{1+D^2}$. For simplify, we will take the Boltzmann constant $k = 1$ in the following calculation.

The entanglement of two qubits can be measured by the concurrence $C$, which is defined as $C = \max[0, 2\max(\sqrt{\lambda_i}) - \sum_{i=1}^{4}\sqrt{\lambda_i}]$ [19], where $\lambda_i$ are the eigenvalues of the matrix $\Gamma = \rho\Theta\rho^*\Theta$, $\rho$ is the density matrix, and $\Theta = \sigma_{1y} \otimes \sigma_{2y}$. Based on the definition of concurrence, the entanglement of two-qubits Heisenberg XX chain with DM interaction at finite temperature can be obtained as [19]:

$$C = \frac{2}{Z}\max[\frac{1}{2}\left|e^{\frac{\beta(J-\delta)}{2}}(1-e^{\beta\delta})\right| - e^{\frac{\beta J}{2}}, 0]. \tag{11}$$

The evolution of the concurrence with respect to $D$ and $J$ for different temperature is shown in Fig.2. It can be seen from the figure that the entanglement decreases gradually with the increase of temperature.

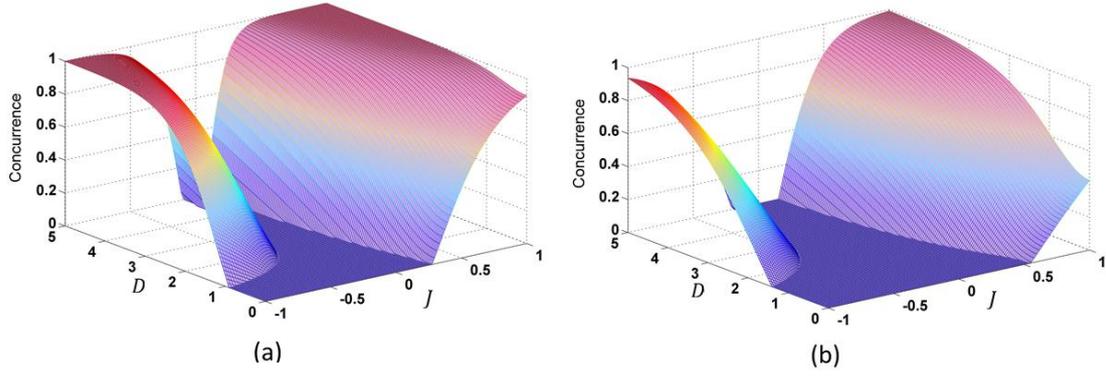

Fig.2: The evolution of the concurrence with $D$ and $J$ for different value of $T$, $T=0.5$ in (a) and $T=1$ in (b).

For the density matrix $\rho$, the state is a pure one when $\text{tr}(\rho^2)=1$, and $\text{tr}(\rho^2)<1$ for the mixed one. Denoting $1-\text{tr}(\rho^2)$ by $\Upsilon$, one can deduce that the bigger the value of $\Upsilon$ is the bigger the mixedness of $\rho$ is. Therefore the value of $\Upsilon$ can be employed to detect the mixedness of qubits states [20, 21]. In addition we can obtain the convexity of the mixedness as (for more detail please see Appendix A):

$$\Upsilon(x\rho_A+(1-x)\rho_B)\geq x\Upsilon(\rho_A)+(1-x)\Upsilon(\rho_B), \quad (12)$$

here $\rho_A$ and $\rho_B$ represent two arbitrary density matrices, $x\rho_A+(1-x)\rho_B$ is the combination of them with $0\leq x\leq 1$, $\Upsilon(\rho)$ stands for the mixedness of the state $\rho$ ($\rho\in\{\rho_A,\rho_B,x\rho_A+(1-x)\rho_B\}$).

Taking advantage of Eq. (6) and the definition of $\Upsilon$, one can acquire:

$$\Upsilon=\frac{4e^{\beta(J+\delta)}[\cosh(\beta J)+2\cosh(\frac{\beta\delta}{2})]}{[e^{\beta(J+\delta)}+e^{\beta J}+2e^{\frac{\beta\delta}{2}}]^2}. \quad (13)$$

Compare with entanglement, the evolution of $\Upsilon$ with respect to $D$ and $J$ for different temperature is shown in Fig.3.

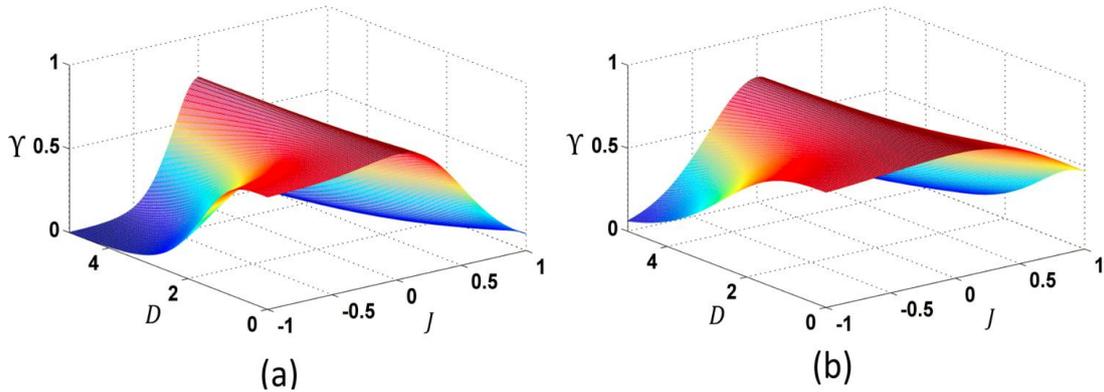

Fig.3: The evolution of $\Upsilon$ with $D$ and $J$ for different value of $T$, $T=0.5$ in (a) and $T=1$ in (b).

Different from entanglement, the mixedness has a positive relation with the temperature. In addition, one can find that the evolution of mixedness is almost contrary to the one of entanglement by comparing Fig.3 with Fig.2. Thus we hold the view that similar to entanglement, the mixedness can also be used as a physical quantity to research the properties of uncertainty relation.

**III. The Effects of Mixedness and Entanglement on Entropic Uncertainty Relation**

In this section, the influence of entanglement and mixedness on the entropic uncertainty will be investigated. Here we take:

$$R = \sigma_x \quad S = \sigma_z, \tag{14}$$

We have $\log_2(1/c) = 1$ by a simple calculation (for more detail please see Appendix B). Taking advantage of Eqs. (6), (14) and the definition of the conditional von Neumann entropy, one can obtain(for more detail please see Appendix B):

$$H(R|B) = -\frac{\Lambda_1 \log_2\left[\frac{\Lambda_1}{2(\Lambda_1+\Lambda_2)}\right] + \Lambda_2 \log_2\left[\frac{\Lambda_2}{2(\Lambda_1+\Lambda_2)}\right]}{(\Lambda_1+\Lambda_2)} - 1 \quad, \tag{15}$$

$$H(S|B) = -\frac{\Delta_1 \log_2\left[\frac{\Delta_1}{2(\Delta_1+\Delta_2)}\right] + \Delta_2 \log_2\left[\frac{\Delta_2}{2(\Delta_1+\Delta_2)}\right]}{(\Delta_1+\Delta_2)} - 1 \quad, \tag{16}$$

$$H(A|B) = \frac{\log_2 O_1}{O_1} + \frac{2\log_2 O_2}{O_2} + \frac{O_3 \log_2 O_2}{O_2} - \frac{O_3 \log_2 O_3}{O_2} - 1 \quad, \tag{17}$$

in which $\Lambda_1 = \exp(\beta J) + \exp(\delta\beta/2)$, $\Lambda_2 = \exp(\delta\beta/2) + \exp(\beta(J + \delta))$, $\Delta_1 = 1$, $\Delta_2 = \exp(\beta J)\cosh(\delta\beta/2)$, $O_1 = 1 + \exp(\delta\beta) + 2\exp(-J\beta + \beta\delta/2)$, $O_2 = Z\exp(\beta J/2)$ and $O_3 = \exp(J\beta + \beta\delta/2)$.

In order to investigate the properties of the entropy uncertainty relation based on the Heisenberg model, we define:

$$W = \log_2 \frac{1}{c} + H(A|B) \quad, \tag{18}$$

$$U = H(R|B) + H(S|B) / (\log_2 \frac{1}{c} + H(A|B)) \quad, \tag{19}$$

$$V = H(R|B) + H(S|B) - \log_2 \frac{1}{c} - H(A|B) \quad. \tag{20}$$

The lower bound of uncertainty $W$ can be used to measure the quality of an uncertain relationship. The smaller the value of $W$ is the better the quality of the uncertainty is, and what's more, the measurement result of R and S can be correctly forecasted if the lower bound $W$ equals 0[22]. On behalf of the ratio and difference of the left to right side of the uncertainty, $U$ and $V$ can be employed to measure the tightness of the uncertain relations. The evolution of $W$, $U$ and $V$ with respect to the temperature in the Heisenberg model is shown in the following figure.

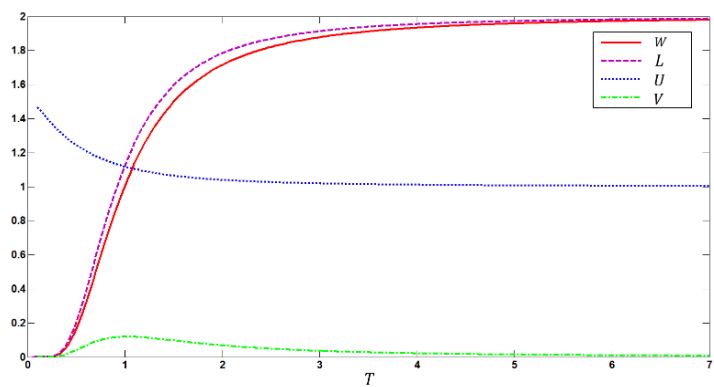

Fig.4: The evolution of $W$, $U$ and $V$ with T, $L$ stands for the left hand of the uncertainty. Here we take $D = 1$ and $J = 1$

From the Fig.4, one can see that the lower bound and the left side of the uncertainty is getting larger with the increase of temperature. In addition, the lower bound W has an negative relationship with the value of $U$ and $V$ on the condition that the value of the T is relatively high, which means the larger the lower bound is the better the tightness is at the case of higher temperature.

In the following, we mainly focus on the effect of the mixedness and the entanglement on the entropic uncertain relation from the perspectives of $W$, $U$ and $V$.

### A. Measuring the quality of the entropy uncertainty by $W$

In this section, we mainly focus on the lower bound of the entropic uncertainty. By means of Eqs. (17) and (18), one can acquire the evolution of the lower bound with respect to $D$ and $J$ for different temperature in Fig.5.

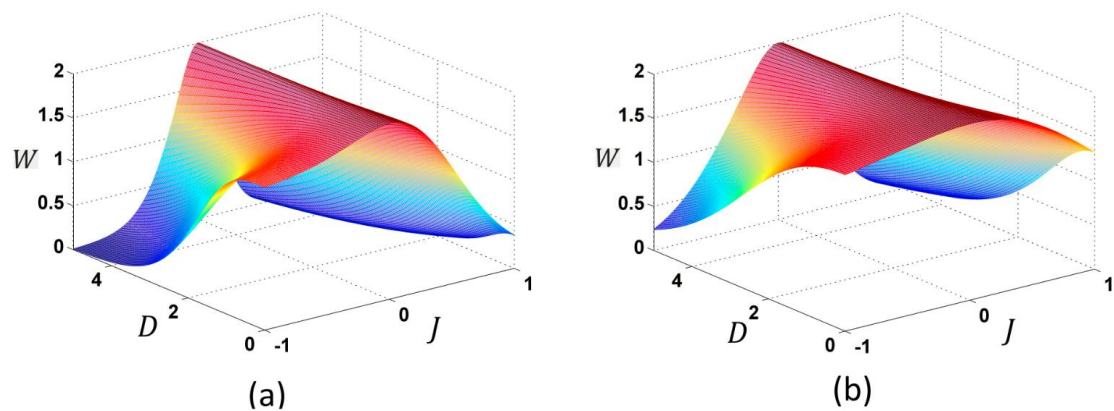

Fig.5: The evolution of $W$ with $D$ and $J$ for different value of $T$, $T = 0.5$ in (a) and $T = 1$ in (b).

As shown in Fig.5, the lower bound is improved with the increase of the temperature and we have $W = 0$ for a particular $(D, J)$ on the condition that the temperature is very small. Comparing Fig.5 with Fig.2, one can see that the stronger the entanglement is, the smaller the lower bound of the uncertainty is. In other words, the entanglement between the measured subsystem and memory one can reduce the uncertainty of measurement results. Meanwhile, we can deduce that the lower bound can be improved with the increase of mixedness by making a comparison between Fig.5 and Fig.3. In addition we find the Fig.5 share a similar structure with the Fig.3, that is to say the lower bound has a close relationship with the mixedness. The evolutions of $W, C$ and $\Upsilon$ can be acquired by making $D = 1$ and $J = 1$ in Fig. 6.

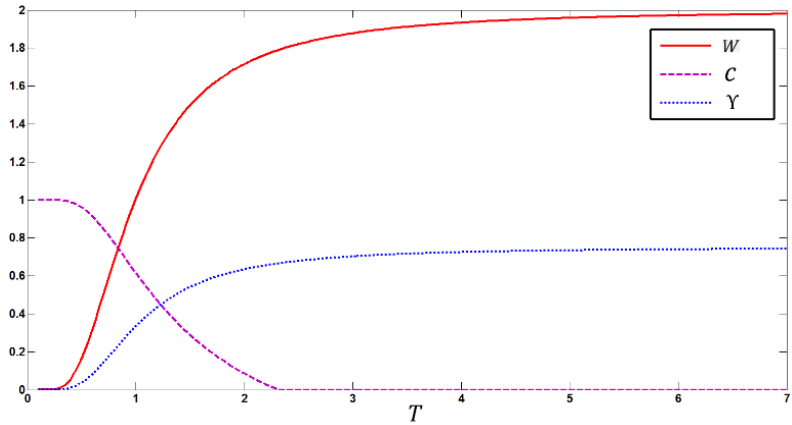

Fig.6: the evolution of $W$, $C$ and $\Upsilon$ with $T$, here we take $D = 1$ and $J = 1$.

As shown above, different from the entanglement, the mixedness of the system share a positive and single-valued function relationship with the lower bound. Therefore we hold the view that the mixedness has a closer link with the lower bound than the entanglement.

### B. Measuring the tightness of the entropy uncertainty by $U$

In this section we mainly investigate the radio tightness of the entropy uncertainty by $U$, a physical quantity defined as the ratio of the left to right side of the uncertainty. Taking advantage of Eqs. (15), (16), (17) and (19), one can obtain the evolution of the $U$ with respect to $D$ and $J$ for different temperature in Fig.7.

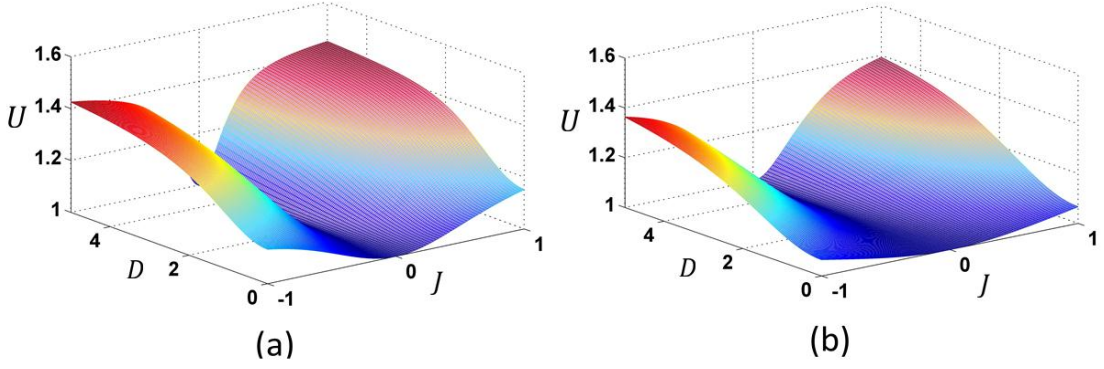

Fig.7: The evolution of $U$ with $D$ and $J$ for different value of $T$, $T = 0.5$ in (a) and $T = 1$ in (b).

From the figure we can see that the value of $U$ will be reduced with the increasing of the temperature. That is to say, the higher temperature will improve the tightness of the uncertain relations. Comparing Fig.7 with Fig.2 one can see that the stronger the entanglement is, the worse the tightness of the uncertainty is. In other words, the entanglement between the measured subsystems and memory one can destroy the tightness of the uncertainty. Meanwhile we can find that the evolution of $U$ has an opposite structure with the one of $\Upsilon$ by making a comparison between Fig.7 and Fig.3, which means the tightness of the uncertainty will be improved with the increasing of the mixedness. Taking $D = 1$ and $J = 1$, one can obtain the evolution of $U$, $C$ and $\Upsilon$ along with $T$, as shown in Fig.8.

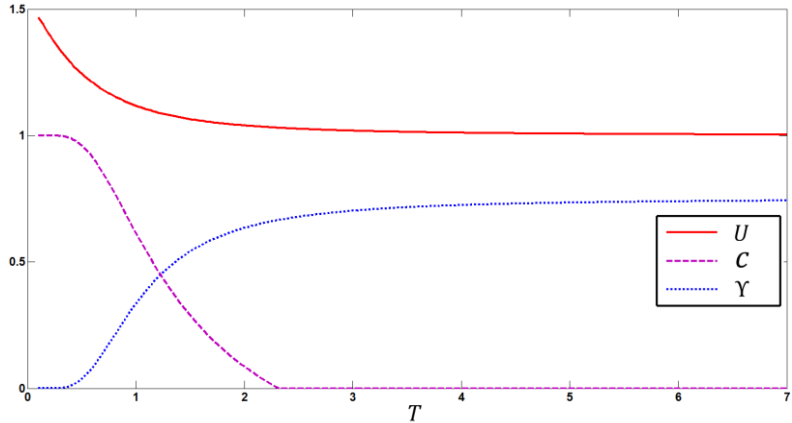

Fig.8: The evolution of $U$, $C$ and $\Upsilon$ with $T$, here we take $D = 1$ and $J = 1$.

Similar to lower bound, the tightness keep a single-valued relationship with the mixedness of the state. Therefore the conclusion that the mixedness can reflect the tightness of the uncertainty better than the entanglement can be obtained.

### C. Measuring the tightness of the entropy uncertainty by $V$

Similar to $U$, the value of $V$, a physical quantity defined as the difference of the left to right side of the uncertainty, will be employed to measuring the tightness of the entropy uncertainty in this section, The evolutions of the $V, C$ and $\Upsilon$ with respect to $T$ can be obtained by taking advantage of Eqs. (15), (16), (17) and (20), as shown in Fig.9.

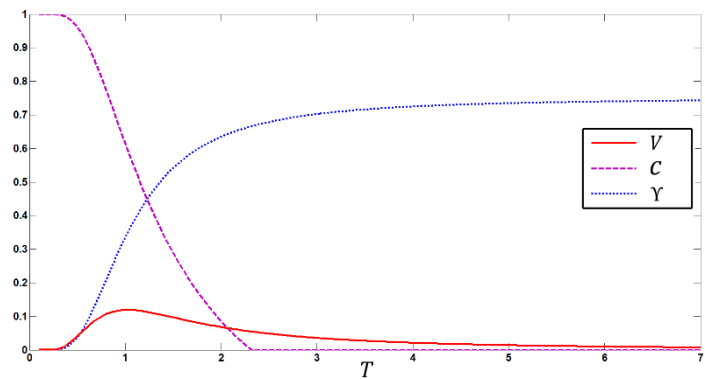

Fig.9: The evolution of $V$, $C$ and $\Upsilon$ with $T$ here we take $D = 1$ and $J = 1$.

As we can see from Fig.9, different from $W$ and $U$, the value of $V$ has no monotonic relationship with $\Upsilon$. However we can deduce that the tightness $V$, which still keeps a single-valued relationship with $\Upsilon$, can be expressed as a function of the mixedness. Let $J$ equal to a specific value, thus the tightness $V$ is a function with respect to $(D, T)$, so are $\Upsilon$ and $D$. Basing on the above analysis, we can obtain the evolution of $V$ with respect to $(\Upsilon, D)$. In addition, we find that the evolution has nothing to with the value of $J$ and is only determined by the sign of $J$. In a word, the tightness of the uncertainty in Heisenberg model can be expressed as the function of the strength of the DM interaction and the mixedness of the system, what's more the functional form does not change with temperature and only has a relationship with magnetic properties. The evolution is shown in Fig.10.

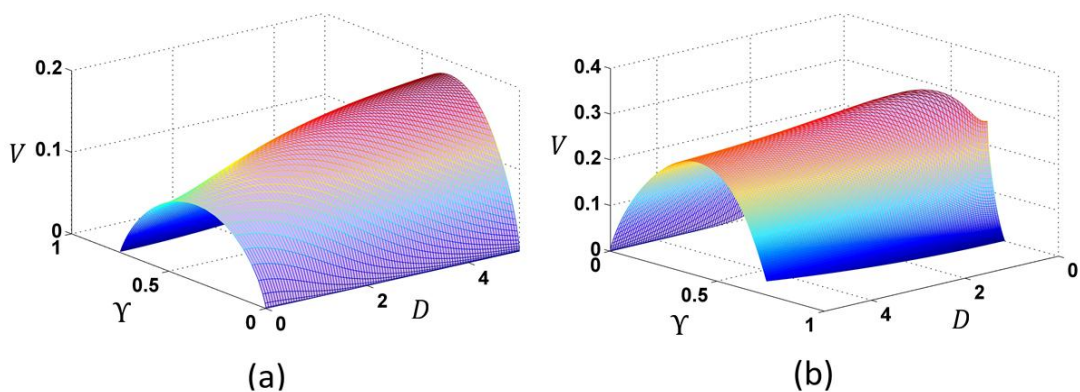

Fig.10: The evolution of $V$ with respect to $\Upsilon$ and $D$, $J > 0$ in (a), $J < 0$ in (b).

As shown in Fig.10, the value of V is equal to zero when the value of $\Upsilon$ reach the minimum,

which means the uncertain inequality becomes uncertain equality at the case that the system is in pure state. At the same time, when the mixedness is determined, the tightness will be reduced with the increasing of the strength of DM interaction on the condition that the system is in the antiferromagnetic coupling，while the situation is just the opposite when the system is in the ferromagnetic coupling.

## IV. Conclusions

In conclusion, we have investigated the effect of the mixedness and the entanglement on the properties of the entropic uncertainty in Heisenberg model with DM interaction. First of all, the uncertainty of measurement results will be reduced by the entanglement between the measured subsystem and memory one while be improved by mixedness of the system. Meanwhile, the entanglement can destroy the tightness of the uncertainty, while the tightness will be improved with the increasing of the mixedness. In addition , we find that the tightness of the uncertainty in Heisenberg model can be expressed as the function of magnetic properties, the strength of the DM interaction and the mixedness of the state, what's more the functional form do not change with temperature. At the same time the uncertain inequality becomes uncertain equality on the condition that the mixedness reaches the minimum value. For a given mixedness, the tightness will be reduced with the increasing of the strength of DM interaction at the case that the system is in the antiferromagnetic coupling，and the situation is just the opposite when the system is in the ferromagnetic coupling.

**Acknowledgments**

This work is supported by the National Natural Science Foundation of China (Grant No. 11574022 and 11174024) and the Open Project Program of State Key Laboratory of Low-Dimensional Quantum Physics (Tsinghua University) grants Nos. KF201407 and supported by the exploratory study of State Key Laboratory of Software Development Environment grants Nos. SKLSDE-2015ZX-10.

## Appendix A

For $d$ dimensional density matrix space, the inner product is defined as:

$$\langle \rho, \sigma \rangle = \text{tr}(\rho^+ \sigma) \tag{A1}$$

And then we can obtain a set of linearly independent basis for the space

$$\{I，\Pi_1, \Pi_2, \dots \Pi_{d-1}\} \tag{A2}$$

where I stands for the identical operator, $\text{tr}(\Pi_i) = 0$, and $\langle \Pi_i, \Pi_j \rangle = \delta_{ij}$. Thus two arbitrary density matrixes in the Bloch sphere can be expressed as

$$\rho_A = \frac{1}{d}(I + \sum_{i=1}^{d-1} p_i^A \Pi_i) \tag{A3}$$

$$\rho_B = \frac{1}{d}(I + \sum_{i=1}^{d-1} p_i^B \Pi_i) \tag{A4}$$

Here $0 \leq \sum_{i=1}^{d-1} p_i^{A^2} \leq 1$ and $\sum_{i=1}^{d-1} p_i^{A^2} = 1$ corresponds to the pure state. In addition, taking advantage of Eqs.(A3) and (A4), one can obtain the combination of $\rho_A$ and $\rho_B$:

$$x\rho_A + (1-x)\rho_B = \frac{1}{d}\{I + \sum_{i=1}^{d-1}[xp_i^A + (1-x)p_i^B\Pi_i]\} \tag{A5}$$

with $0 \leq x \leq 1$. Making use of the Hermitian of density matrix and the definition of the inner product, one can obtain:

$$\mathrm{tr}(\rho_A^2) = \mathrm{tr}(\rho_A^+ \rho_A) = \frac{1}{d^2}(d + \sum_{i=1}^{d-1} p_i^{A^2}) \tag{A6}$$

$$\mathrm{tr}(\rho_B^2) = \mathrm{tr}(\rho_B^+ \rho_B) = \frac{1}{d^2}(d + \sum_{i=1}^{d-1} p_i^{B^2}) \tag{A7}$$

$$\mathrm{tr}([x\rho_A + (1-x)\rho_B]^2) = \frac{1}{d^2}(d + \sum_{i=1}^{d-1}[xp_i^A + (1-x)p_i^B]^2) \tag{A8}$$

and then we have

$$\mathrm{tr}([x\rho_A + (1-x)\rho_B]^2) \leq x\mathrm{tr}(\rho_A^2) + (1-x)\mathrm{tr}(\rho_B^2) \tag{A9}$$

where the inequalities $p_i^{A^2} + p_i^{B^2} \geq 2p_i^A p_i^B$ has been used. Thus we can deduce the convexity of the mixedness

$$\Upsilon(x\rho_A + (1-x)\rho_B) \geq x\Upsilon(\rho_A) + (1-x)\Upsilon(\rho_B) \tag{A10}$$

In addition, according to Eqs. (A6) and (A7), one can acquire that the mixedness reach the maximum value $1 - \frac{1}{d}$ on the condition that mold of Bloch vector $\sum_{i=1}^{d-1} p_i^2 = 0$, and at this moment, the density matrix has the following form:

$$\rho = \frac{1}{d}I \tag{A11}$$

**Appendix B**

As mentioned above, $H(Y|B)$ is the conditional von Neumann entropy of the post measurement state $\rho_{YB} = \sum_Y (|\Psi_Y\rangle\langle\Psi_Y| \otimes I)\rho_{AB}(|\Psi_Y\rangle\langle\Psi_Y| \otimes I)$ with $|\Psi_Y\rangle$ being the eigenvectors of $Y$, $Y \in (R, S)$. Taking advantage of Eq. (14), one can obtain

$$|\Psi_{R1}\rangle = \frac{1}{\sqrt{2}}\begin{pmatrix}-1\\1\end{pmatrix}, \quad |\Psi_{R2}\rangle = \frac{1}{\sqrt{2}}\begin{pmatrix}1\\1\end{pmatrix} \tag{B1}$$

$$|\Phi_{S1}\rangle = \begin{pmatrix}1\\0\end{pmatrix}, \quad |\Phi_{S2}\rangle = \begin{pmatrix}0\\1\end{pmatrix} \tag{B2}$$

in which $|\Psi_{Ri}\rangle$ and $|\Psi_{Si}\rangle$ represent the eigenstates of $R$ and $S$ respectively $i \in (1,2)$. And then according to the definition of $\rho_{RB}$, $\rho_{SB}$ and $\rho_B$ we can acquire them in the Heisenberg Model with DM Interaction, which read as

$$\rho_{RB} = \begin{pmatrix} \frac{1}{4} & 0 & 0 & \frac{e^{\beta J+i\theta}(1-e^{\beta\delta})}{4(\Lambda_1+\Lambda_2)} \\ 0 & \frac{1}{4} & \frac{e^{\beta J-i\theta}(1-e^{\beta\delta})}{4(\Lambda_1+\Lambda_2)} & 0 \\ 0 & \frac{e^{\beta J+i\theta}(1-e^{\beta\delta})}{4(\Lambda_1+\Lambda_2)} & \frac{1}{4} & 0 \\ \frac{e^{\beta J-i\theta}(1-e^{\beta\delta})}{4(\Lambda_1+\Lambda_2)} & 0 & 0 & \frac{1}{4} \end{pmatrix} \quad (B3)$$

$$\rho_{SB} = \begin{pmatrix} \frac{1}{2+2\Delta_2} & 0 & 0 & 0 \\ 0 & \frac{1}{2}-\frac{1}{2+2\Delta_2} & 0 & 0 \\ 0 & 0 & \frac{1}{2}-\frac{1}{2+2\Delta_2} & 0 \\ 0 & 0 & 0 & \frac{1}{2+2\Delta_2} \end{pmatrix} \quad (B4)$$

$$\rho_B = \begin{pmatrix} \frac{1}{2} & 0 \\ 0 & \frac{1}{2} \end{pmatrix} \quad (B5)$$

Therefore, we can obtain

$$H(\rho_{RB}) = \sum -\lambda_{rbi}\log_2(\lambda_{rbi}) = -\frac{\Lambda_1\log_2\left[\frac{\Lambda_1}{2(\Lambda_1+\Lambda_2)}\right]+\Lambda_2\log_2\left[\frac{\Lambda_2}{2(\Lambda_1+\Lambda_2)}\right]}{(\Lambda_1+\Lambda_2)} \quad (B6)$$

$$H(\rho_{SB}) = \sum -\lambda_{sbi}\log_2(\lambda_{sbi}) = -\frac{\Delta_1\log_2\left[\frac{\Delta_1}{2(\Delta_1+\Delta_2)}\right]+\Delta_2\log_2\left[\frac{\Delta_2}{2(\Delta_1+\Delta_2)}\right]}{(\Delta_1+\Delta_2)} \quad (B7)$$

$$H(\rho_{AB}) = \sum -\lambda_{abi}\log_2(\lambda_{abi}) = \frac{\log_2 O_1}{O_1} + \frac{2\log_2 O_2}{O_2} + \frac{O_3\log_2 O_2}{O_2} - \frac{O_3\log_2 O_3}{O_2} \quad (B8)$$

$$H(\rho_B) = \sum -\lambda_{bi}\log_2(\lambda_{bi}) = 1 \quad (B9)$$

where $\lambda_{rbi}$, $\lambda_{sbi}$, $\lambda_{abi}$ and $\lambda_{bi}$ are the eigenvalues of $\rho_{RB}$, $\rho_{SB}$, $\rho(T)$ and $\rho_B$. Making use of the definition of the conditional von Neumann entropy, we can deduce Eqs. (15), (16) and (17).

In addition putting Eqs. (B1) and (B2) into $c = \max_{r,s}|\langle\Psi_r|\Phi_s\rangle|^2$, we can obtain $c = 1/2$. Therefore we have

$$\log_2\frac{1}{c} = 1 \quad (B10)$$